# Electric-Field-Modulated Nonvolatile Resistance Switching in VO$_2$/PMN-PT(111) Heterostructures

Bowen Zhi[†], Guanyin Gao[*,†], Haoran Xu[†], Feng Chen[‡], Xuelian Tan[†], Pingfan Chen[†], Lingfei Wang[†] and Wenbin Wu[*, †, ‡]

[†]Hefei National Laboratory for Physical Sciences at Microscale, University of Science and Technology of China, Hefei 230026, People's Republic of China

[‡]High Magnetic Field Laboratory, Chinese Academy of Science (CAS), Hefei 230071, People's Republic of China

ABSTRACT: The electric-field-modulated resistance switching in VO$_2$ thin films grown on piezoelectric (111)-0.68Pb(Mg$_{1/3}$Nb$_{2/3}$)O$_3$-0.32PbTiO$_3$ (PMN-PT) substrates has been investigated. Large relative change in resistance (10.7%) was observed in VO$_2$/PMN-PT(111) hererostructures at room temperature. For a substrate with a given polarization direction, stable resistive states of VO$_2$ films can be realized even when the applied electric fields are removed from the heterostructures. By sweeping electric fields across the heterostructure appropriately, multiple resistive states can be achieved. These stable resistive states result from the different stable remnant strain states of substrate, which is related to the rearrangements of ferroelectric domain structures in PMN-PT(111) substrate. The resistance switching tuned by electric field in our work may have potential applications for novel electronic devices.

1. Introduction

Nonvolatile random access memory (RAM) has attracted much attention in the last decades because of its tremendous potential application in the semiconductor industry. In this field, how to improve the ability on reading and writing nonvolatile information with lower energy consumption remains an important and tantalizing issue.[1-3] Recently, switching of various physical parameters between different states by controlling strain becomes a growing concern[4-14] and has been achieved in many kinds of magnetic materials such as metal, manganite and alloys.[4,8,9,14] One of the methods is to integrate functional films with ferroelectric substrate and then tuning the electric or magnetic states of the films via the inverse piezoelectric effect by applying an appropriate electric field across the ferroelectric substrates.[10] Tuning the magnetic states by electric field, which is also known as the converse magnetoeletric effect, plays a dominating role in related research work.[8,9,14] On the other hand, tuning the resistance states, especially the nonvolatile resistance switching via strain, is a relatively new field. Traditional resistance switching in a metal-insulator-metal structure has a mechanism which is generally explained with filamentary conduction,[3,15] but nonvolatile resistance switching via strain has a different mechanism and provides a potential strategy for designing prototype devices in information storage industry.[6,11]

A challenge in nonvolatile switching is how to retain the induced state even after removing the electric field. Recently, two reversible and stable remnant strain states were achieved in PMN-PT single crystals.[13,14] Moreover, it is reported that there exists an intermediate state with most of the ferroelectric domains lying in the in-plane direction.[6] These results indicate a potential method for nonvolatile switching. Actually, based on this principle, Yang *et al.* has observed three remnant strain states and manipulated three resistance states in $La_{2/3}Sr_{1/3}MnO_3/0.7Pb(Mg_{2/3}Nb_{1/3})O_3$-$0.3PbTiO_3$ heterostructures.[11] However, the relative change in resistance is too small (1%) for practical application. Consequently, it is of considerable interest to find other materials which have more remarkable relative resistance change. Vanadium dioxide ($VO_2$) has attracted much attention in the last decades,[16-22] not only because of its metal-insulator phase transition accompanied by an abrupt change in resistivity and near-infrared optical properties,[16,20] but also because of the many technological applications such as optical switching and smart window.[21,22] Up to now, $VO_2$ films have been fabricated on various kinds of substrates such as $TiO_2$ and sapphire.[23,24] The metal-insulator phase transition of $VO_2$ films is very sensitive to their strain states and it is difficult to continuously tune their strain state. Thus, it will be fantastic if a structure could be realized in which the strain state and resistivity of $VO_2$ films can be tuned continuously.

In this work, $VO_2$/PMN-PT(111) heterostructures were fabricated to investigate their nonvolatile switching through ferroelectric polarization. We observe that the resistance can be switched effectively between low resistance and high resistance states by adjusting the electric fields across the PMN-PT substrate, and interestingly the high resistance state can be retained even after decreasing the electric fields to zero in the unipolar sweeping process. A relative change in the resistance can be up to 10.7% at 300 K near the coercive electric field. Benefiting from this unique behavior, at least four stable resistance states can be generated readily.

2. Experimental Section

VO$_2$ films were grown on (111)-0.68Pb(Mg$_{1/3}$Nb$_{2/3}$)O$_3$-0.32PbTiO$_3$ (PMN-PT) commercial single crystal substrates (CMT Corporation) by pulsed laser deposition. PMN-PT single crystal is a relaxor ferroelectric, in which monoclinic, rhombohedral, and tetragonal phases coexist. Consequently, this material shows excellent ferroelectric and piezoelectric properties. The V$_2$O$_5$ ceramic target was prepared by standard solid state reactions. The laser energy density was kept at about 2 J/cm$^2$ and the repetition rate was 5 Hz. The laser ablation was performed under an oxygen pressure ($P_O$) fixed at 0.1 mTorr, with the substrate temperature at 500 $^o$C. The film thicknesses were determined to be about 150 nm by scanning electron microscopy (SEM). Film surface morphologies were analyzed by atomic force microscopy (AFM, Vecco, MultiMode V). The structure of the films was characterized by High-resolution X-ray diffraction (XRD) using Cu K$\alpha_1$ ($\lambda$ = 1.5406 Å) radiation (PANalytical X'Pert Pro). Four parallel Pt top electrodes (2.6 mm × 0.7 mm) were prepared using a shadow mask. The resistance measurements were performed using the standard four point method on a superconducting quantum interference device (SQUID) magnetometer (MPMS, Quantum Design). The distance between two adjacent electrodes was 0.7 mm. The resistivity of VO$_2$ films was 0.64 Ω·cm at 200 K and 3.2 × 10$^{-4}$ Ω·cm at 340 K, which was much smaller than that of PMN-PT substrates (~10$^{10}$ Ω·cm). Thus, the contribution to the resistance measurement from PMN-PT substrates could be neglected. The in-plane strain of the sample was measured by mounting a strain gauge on the sample surface.[11-13]

3. Result and Discussion

   Tetragonal phase of $VO_2$ has a quadruple symmetry and PMN-PT(111) has a 3-fold symmetry. The difference of symmetry makes it difficult to deposit $VO_2$ films epitaxially on PMN-PT(111) substrate. However, $VO_2$ thin film can be grown on PMN-PT(111) substrate using domain matching epitaxy.[23] Figure 1(a) shows XRD pattern of $VO_2$/PMN-PT(111) heterostructure in the unpolarized state. The $VO_2$ film is (0k0)-oriented and no other orientations are found. Besides, the (020) peak for $VO_2$ film is the same as the peak of $VO_2$ bulk, indicating the initial strain due to lattice mismatch has been relaxed. The inset of Figure 1(a) presents an AFM image obtained from the $VO_2$ film. The image shows the $VO_2$ film has an oriented polycrystalline structure with elongated, acicular grains. These grains have three in-plane orientations and the angle between two adjacent grains is about 120°, indicating the $VO_2$ film exhibits textured growth on PMN-PT(111) substrate. According to the results in AFM image, there are three possible arrangements for $VO_2$ unit cells on PMN-PT(111) substrate. Figure 1(b) shows the schematic of in-plane arrangements for rutile $VO_2$ unit cells on PMN-PT(111) substrate. The in-plane orientation relationship between $VO_2$(020) film and PMN-PT(111) substrate can be written as $[200]_{VO2}$ ∥ $[\bar{2}11]_{PMN-PT}$ and $[002]_{VO2}$ ∥ $[01\bar{1}]_{PMN-PT}$. The mismatch is 7.6% along $VO_2$ [002] and -0.2% along $VO_2$ [200].

  In situ XRD linear scans were measured in order to investigate the structural evolution of $VO_2$/PMN-PT(111) heterostructures from unpolarized to polarized states. The inset of Fig. 2(b) shows the schematic of the $VO_2$/PMN-PT(111) structure with electrical field configuration, where Pt film was deposited on the back of the PMN-PT substrate as bottom electrode. A DC electric field of 6 kV/cm across the $VO_2$/PMN-PT(111) heterostructure was applied for about twenty minutes to ensure the PMN-PT substrate was absolutely polarized. As shown in Fig. 2(a), after the PMN-PT was polarized, both the VO$_2$(020) and the PMN-PT(111) reflection peak shifted to lower Bragg angle, meaning that the out-of-plane lattice parameter increases. It is found that the out-of-plane lattice parameters of PMN-PT(111) and VO$_2$ were enlarged by 0.10% and 0.06%, respectively. Therefore, the induced strain from ferroelectric polarization of the PMN-PT(111) substrate can be transferred to the VO$_2$ film, but strain relaxation should also exist due to structural defects in VO$_2$ film. Temperature-dependent resistance curves of VO$_2$ film in the unpolarized and polarized states were also measured to observe the effect of ferroelectric polarization on transport behaviors, as shown in Fig. 2(b). The metal-insulator transition temperature is determined from the Gaussian fit of the $d\ln R/dT$ curves. The metal-insulator transition temperature decreases by 3 K after polarizing the PMN-PT substrate. It has been reported that the uniaxial strain along the c-axis in VO$_2$ thin films can strongly affect the metal-insulator transition temperature.[24,25] The metal-insulator transition temperature will decrease when the lattice is compressed along the c-axis and increase when the lattice is expanded along the c-axis. In our case, after the PMN-PT substrate was polarized, the in-plane lattice of PMN-PT was compressed and accordingly the out-of-plane lattice of PMN-PT was elongated. Considering our films are (0k0)-oriented, the c-axis of films should lie in plane. Therefore, the VO$_2$ film will suffer an enhanced in-plane compressive strain along c-axis due to the contraction of in-plane lattice of PMN-PT substrate, resulting in a decreased transition temperature.

Figure 3 shows resistance of the VO$_2$ film as a function of sweeping electric fields at different temperature. All the resistance responses under bipolar sweeping of electric fields (in this case, electric fields exceed the coercive field of PMN-PT) show symmetrical butterfly-like shapes (filled squares, solid line), indicating the modulation of the resistance of the VO$_2$ film is mainly due to inverse piezoelectric effect (strain effect), whereas the ferroelectric field effect is negligible.[12,26] Another important effect is that oxygen deficiencies may be induced by electric field.[27,28] In our case, if oxygen deficiencies were induced during the process, an asymmetrical *R-E* curve will be observed because the density of carriers will be unequal when electric fields with opposite directions are applied. Thus, the formation of oxygen deficiencies, if any, will not play an important role in our case.

As temperature increases, a lower electric field is needed to achieve the peak of resistance, corresponding to the decreasing coercive field of the PMN-PT(111) substrate with decreasing temperature. On the other hand, when cycling the electric field below the coercive field, a hysteresis loop can be achieved (open squares, solid line). There exist two stable resistance states at E = 0 due to the different remnant strain. It can be seen that the in-plane strain also has a linear response to the electric field when the electric field increases from negative maximum to 0 kV/cm or decreases from positive maximum to 0 kV/cm, which is similar with the behaviors observed when using PMN-PT(001) and (011) substrates.[6,10] Besides, the in-plane strain response in our case is smaller than that in PMN-PT(001) and (011) substrates, which we attribute to the smaller electromechanical coupling factor along (111) direction in PMN-PT substrate.

It is reported that the change of resistance is related to the evolution of the ferroelectric domains of PMN-PT substrate. Wu *et al.* demonstrated that there exists a metastable state during the polarization reversal at the vicinity of the coercive field due to non-180° polarization reorientation.[13,14] For the PMN-PT(111) substrate, non-180° polarization reorientation may also happen during the process of sweeping the electric field. Thus, we performed *in situ* X-ray diffraction measurements to gain insight into the structure evolution of the VO$_2$/PMN-PT heterostructure when sweeping the unipolar electric field at 300 K. The electric filed was first swept from -6 to 2 kV/cm and then removed. As shown in Fig. 4(a), the PMN-PT(111) reflection and VO$_2$(020) reflection does not change when the electric field increase from -6 to 0 kV/cm, indicating that the remnant strain state has no change. This behavior is corresponding to the resistance response when sweeping the electric field from the negative maximum to zero shown in Fig. 3. When the electric field increase from 0 to 1.6 kV/cm, the PMN-PT(111) reflection splits from one peak to two peaks. This suggests that the ferroelectric domains of PMN-PT rotate as increasing the electric field. It is speculated that the ferroelectric domains may rotate from $[\bar{1}\bar{1}\bar{1}]$ to $[1\bar{1}1]$, $[11\bar{1}]$, $[\bar{1}11]$, $[\bar{1}\bar{1}1]$, $[1\bar{1}\bar{1}]$ or $[\bar{1}1\bar{1}]$ directions and accordingly induce a metastable state.[13] Besides, the VO$_2$(020) reflection also shifts to higher angle, suggesting the in-plane lattice of VO$_2$ film was expanded and the out-of-plane lattice of VO$_2$ film was compressed. When the electric field increases from 1.6 to 1.8 kV/cm, the relative intensity of the two peaks of PMN-PT substrate changes, indicating the variance of reorientation of ferroelectric domains. Structural evolution is also depicted when the electric field is further swept to just below the coercive field (2 kV/cm) and then reduced to 0. Although the PMN-PT(111) reflection changes from two peaks to one peak, the VO$_2$(020) reflection remains almost the same, which corresponds to the resistance response under unipolar electric field. This phenomenon proves that there exist metastable resistance states and the remnant strain state does not change after removing the electric field.

To understand the structural evolution more clearly, the in-plane strain versus electric field at room temperature was also examined, as shown in Fig. 4(b) and 4(c). The curve has a similar pattern with the resistance-electric field loop in Fig. 3, indicating that the resistance change is induced by the evolution of strain. Besides, the in-plane strain becomes the maximum near the coercive field, indicating that ferroelectric domains evolution from a low strain state to a high strain state. This is corresponding to the XRD measurement in Fig. 4(a). On the other hand, as shown in Fig. 4(c), the strain could be maintained when the electric field was removed, suggesting that the ferroelectric domains were at a metastable state.

To determine multiple metastable resistance states at room temperature, different unipolar loops were cycled. Some typical loops are shown in Fig. 5(a). Although the sweeping range of electric field is different, the resistance can keep stable after removing the electric field. Theoretically, infinite metastable states could be achieved if proper electric field was applied in this structure. The difference among these metastable states is related to the proportion of ferroelectric domains that reoriented from $[\bar{1}\bar{1}\bar{1}]$ direction, which leads to different metastable remnant strain states. On the basis of Fig. 5(a), we successfully construct four resistance states by switching appropriate electric field pulses to the structure, as shown in Fig. 5(b). First, by applying a sequence of 2 kV/cm pulses and -2 kV/cm pulses, corresponding binary "0", "3" resistance state can be generated, respectively. The variance between "0" and "3" resistance state is related to the different remnant strain states, which is corresponding to the structure evolution in Fig. 4. By modulating the pulse electric fields precisely, "1" and "2" resistance states can be easily generated. Because of the resolution of XRD, it is difficult to distinguish the delicate difference among the remnant strain states of "1", "2" and "3" state in Fig. 4. In this way, at least four distinguishable resistance states could be obtained. One of the virtues of the $VO_2$/PMN-PT(111) heterostructure is that the relative change in the resistance at room temperature can reach 10.7%. Besides, this structure has a merit of low energy consumption, because the PMN-PT substrate is a dielectric material with resistance higher than $10^9 \Omega$. Thus, the resistance switching under different unipolar electric field could have potential application for designing new electronic devices.

## 4. SUMMARY

In summary, the electric-field-induced resistance switching in $VO_2$ thin films grown on ferroelectric (111)-0.68Pb($Mg_{1/3}Nb_{2/3}$)$O_3$-0.32PbTiO$_3$ (PMN-PT) substrates has been investigated. The resistance can keep unchanged after decreasing the electric fields to zero in the unipolar sweeping of different electric fields across the PMN-PT substrate. A relative change in the resistance can be up to 10.7% at room temperature. Most importantly, multiple distinguishable nonvolatile resistance states have been realized in this structure with the help of stable remnant strain states. The XRD data suggest that the different remnant strain states origin from the reorientation of ferroelectric domains in PMN-PT(111) substrate. The resistance switching tuned by electric field in our work may have potential applications for novel electronic devices.

AUTHOR INFORMATION

**Corresponding Author**

*Electronic mail: ggy@ustc.edu.cn

*Electronic mail: wuwb@ustc.edu.cn

ACKNOWLEDGMENT

This work was supported by the NSF of China (Grant Nos. 11074237 and 11274287), and the National Basic Research Program of China (Grant Nos. 2009CB929502 and 2012CB927402). Also, partial support by the Research Fund for the Doctoral Program of Higher Education of China (Grant No. 20123402120028) is acknowledged.

28. Du, Y.; Pan, H.; Wang, S.; Wu, T.; Feng, Y. P.; Pan, J.; Wee, A.T. S. Symmetrical Negative Differential Resistance Behavior of a Resistive Switching Device ACS Nano 2012, 6, 2517−2523.

Figure Captions

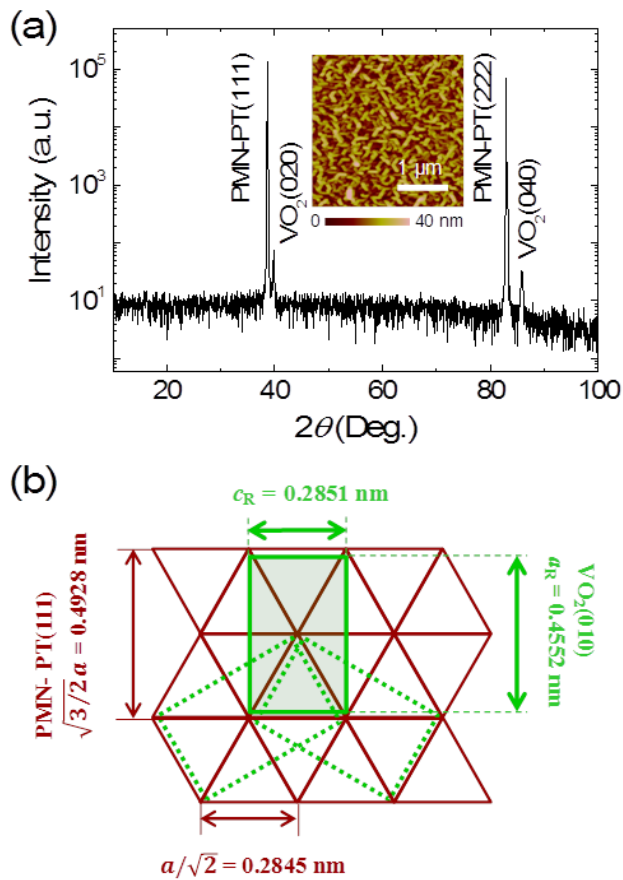

**Figure 1.** (a) XRD pattern of VO$_2$/PMN-PT(111) heterestructure in the unpolarized state. The inset presents AFM image obtained from the VO$_2$ film. (b) Schematic of in-plane arrangements for rutile VO$_2$ unit cells on PMN-PT(111) substrate.

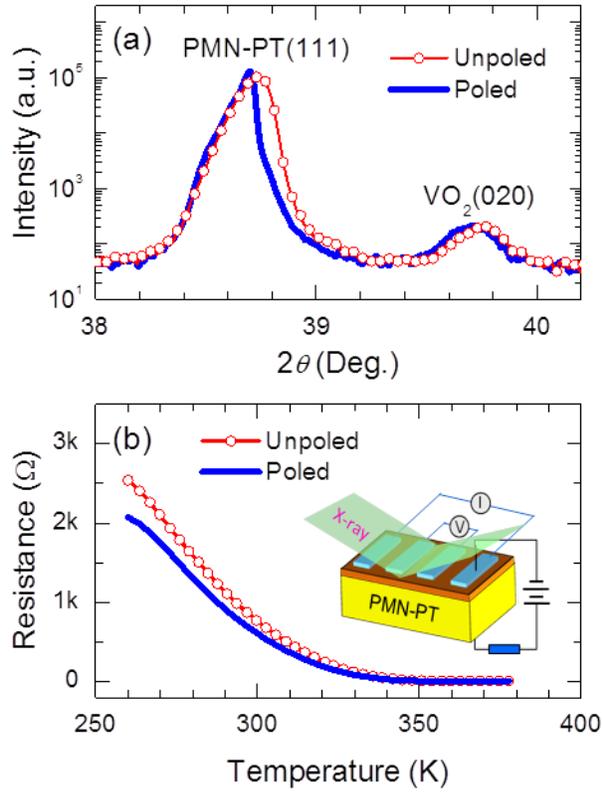

**Figure 2.** (a) In situ XRD linear scans on $VO_2$/PMN-PT(111) heterostructures in the unpolarized and polarized states. (b) Temperature-dependent resistance curves of $VO_2$ film in the unpolarized and polarized states. To be clear, only the heating process is shown here. The inset of (b) shows schematic of the $VO_2$/PMN-PT(111) structure and the electrical field configuration for measurements of the structure and the resistivity.

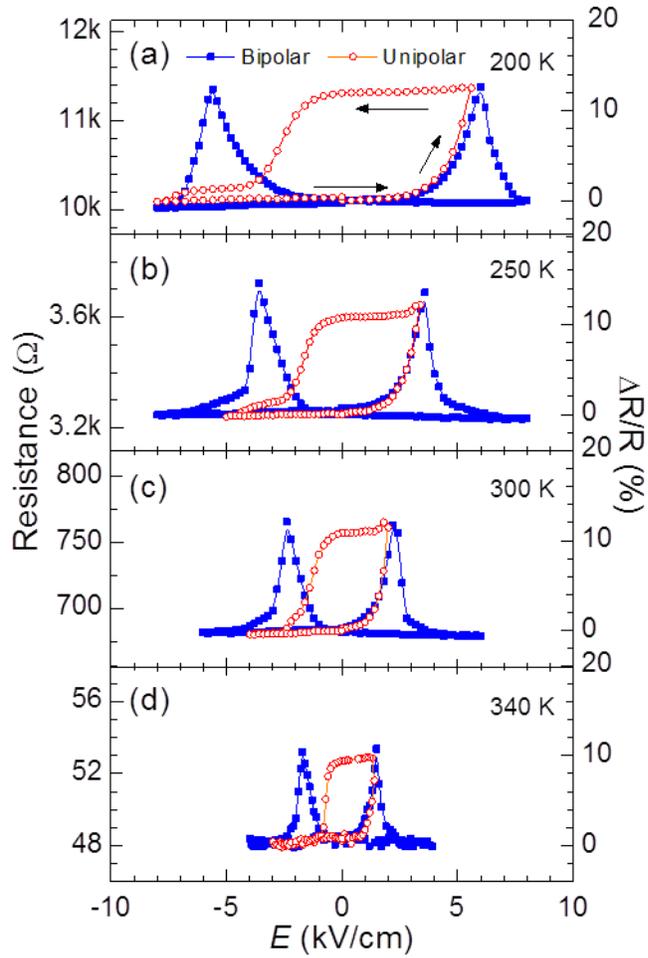

**Figure 3.** Resistance of the VO$_2$ film as a function of electric fields for unipolar and bipolar sweeping at different temperature.

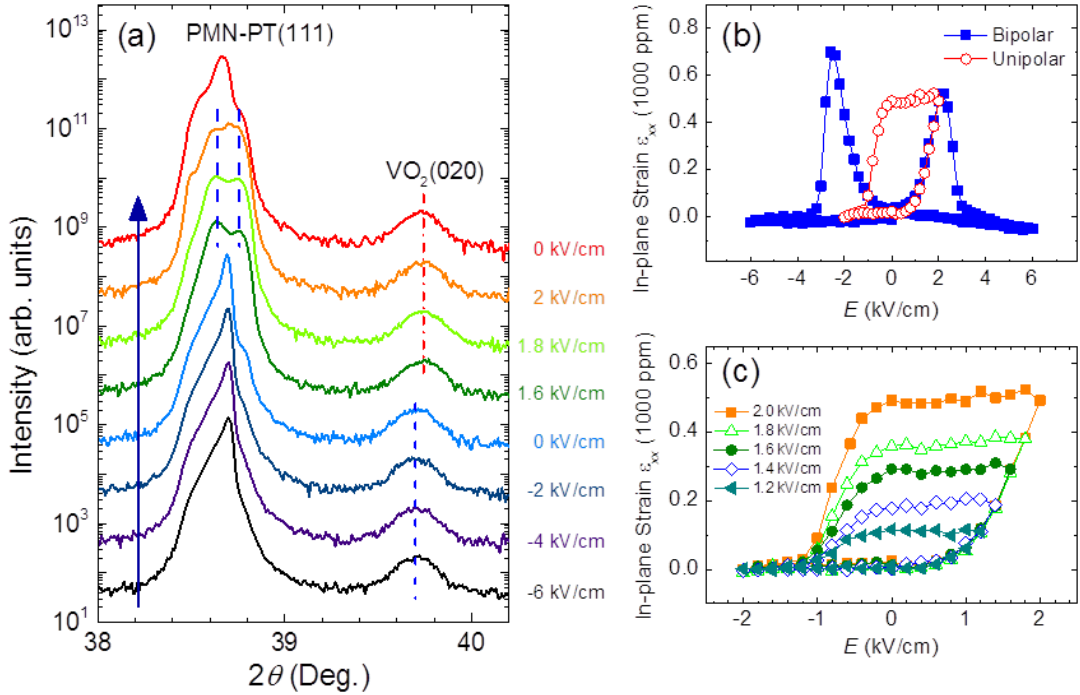

**Figure 4.** (a) *In situ* measurements of the structure evolution of the VO$_2$/PMN-PT heterostructure when applying different electric field at room temperature. (b) In-plane strain versus electric field for unipolar and bipolar sweeping at room temperature. (c) In-plane strain for the VO$_2$/PMN-PT heterostructure versus electric field when cycling the electric field between negative maximum and several positive values at room temperature.

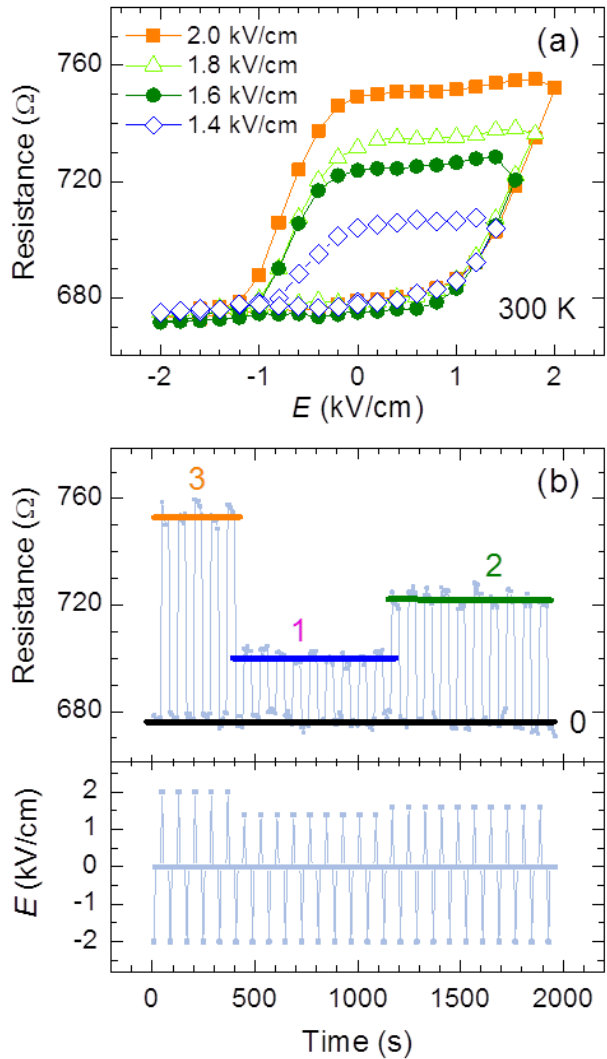

**Figure 5.** (a) Resistance hysteresis loops when cycling the electric field between negative maximum and several positive values at room temperature. (b) Temporal profiles of resistance responses to the electric field pulses. Four nonvolatile resistance states are switched under the electric field pulses.